\magnification1200


\vskip 2cm
\centerline
{\bf  E11 in 11D}
\vskip 1cm
\centerline{ Alexander G. Tumanov and Peter West}
\centerline{Department of Mathematics}
\centerline{King's College, London WC2R 2LS, UK}
\vskip 2cm
\leftline{\sl Abstract}
We construct the non-linear realisation of the semi-direct product of $E_{11}$ and its vector representation in  eleven dimensions and find the dynamical equations it predicts at low levels. These equations are completely determined by the non-linear realisation and when  restricted to contain only the usual fields of supergravity and the  usual space-time we find precisely  the equations of motion of eleven dimensional supergravity. This paper extends  the results announced in  hep-th/1512.01644 and in  particular  it contains the contributions to the equations of motion that involve derivatives with respect to the level one  generalised coordinates.

\vskip2cm
\noindent

\vskip .5cm

\vfill
\eject
{\bf 1 Introduction}
\medskip

Quite some time ago it was conjectured that the low energy effective action for strings and branes is the non-linear realisation of the semi-direct product of $E_{11}$ and its vector ($l_1$) representation,  denoted $E_{11}\otimes _s l_1$ [1,2].  This theory has an infinite number of fields, associated with $E_{11}$,  which live on a generalised spacetime, associated with the vector representation $l_1$. 
\par
The fields obey 
equations of motion that follow from the symmetries of  the non-linear realisation.  Although it was clear from the beginning [1] that the fields at low levels were just those of the maximal supergravity theories the unfamiliar nature of spacetime discouraged the construction of the equations of motion. The earliest attempts often used only the usual coordinates of spacetime and  only  the Lorentz part of the $I_c(E_{11})$ local symmetry.  As a result the full power of the symmetries of the non-linear realisation was not exploited and the results were incomplete.  A more systematic approach was used  to constructing  the equations of motion of the  $E_{11}\otimes_s l_1$ non-linear realisation 
 in eleven [3]  and four [4] dimensions by including  both the higher level generalised coordinates and local symmetries in $I_c(E_{11})$.
These papers did find the  equations of motion of the form fields but found only partial results for the gravity equation.  
\par 
Recently the equations of motion of the non-linear realisation $E_{11}\otimes _s l_1$ were found for all the usual fields in the maximal supergravity fields, including gravity, in five and eleven dimensions [5]. 
The equations of motion in  eleven dimensions were completely determined and in agreement with those of eleven dimensional supergravity when one keeps only the usual supergravity fields and the   usual coordinates of spacetime.  In this paper we will give some of the details of this calculation as well as giving  the terms in the equations of motion that contain derivatives to the level one generalised coordinate. We will also 
complete the variation of the gravity equation under the symmetries of the non-linear realisation to show that it varies in the three form equation. 
\par
In this section we will also review the main features of the non-linear realisation. In section two we will formulate the eleven dimensional theory including the explicit forms of the Cartan forms and generalised vielbein. 
Section three derives the variations of the Cartan form under the symmetries of the non-linear realisation and in particular discusses an important subtilty associated with the fixing of the group element of the non-linear realisation using  its  local symmetry. Using these results in section four we find the equations of motion for the three form and gravity and show that they vary into each other. Finally we discuss some of the consequences of the result in section five. 
\par
To fix the notation,  and as it is still not that well understood,  we recall from previous papers the  main features of the non-linear realisation of $E_{11}\otimes_sl_1$ which is constructed from the group element $g\in E_{11}\otimes_sl_1$ that  can be written as 
$$
g=g_lg_E 
\eqno(1.1)$$
In this equation  $g_E$ is a group element of $E_{11}$ and so can be written in the form 
$g_E=e^{A_{\underline \alpha} R^{\underline \alpha}}$ where the $R^{\underline \alpha}$ are the generators of $E_{11}$ and $A_{\underline\alpha}$ are the fields in the non-realisation. The group element   $g_l$ is formed from the generators of the vector ($l_1$) representation and so has the form $e^{z^A L_A} $ where $z^A$ are the coordinates of the generalised space-time. The fields $A_{\underline\alpha}$ depend on the coordinates $z^A$.  
The non-linear realisation is, by definition, invariant under the transformations 
$$
g\to g_0 g, \ \ \ g_0\in E_{11}\otimes _s l_1,\ \ {\rm as \  well \  as} \ \ \ g\to gh, \ \ \ h\in
I_c(E_{11})
\eqno(1.2)$$
The group element $g_0\in E_{11}$ is a rigid transformation, that is, it is  a constant. The group element $h$ belongs to the Cartan involution invariant subalgebra of $E_{11}$, denoted $I_c(E_{11})$; it is a local
transformation meaning that  it depends on the generalised space-time. 
The action of the Cartan involution can be taken to be 
$I_c(R^{\underline \alpha}) = - R^{-\underline \alpha} $ 
for any root $\alpha$ and so the Cartan involution invariant subalgebra is generated by $R^{\underline \alpha} - R^{-\underline \alpha} $.

As the generators in $g_l$ form a representation of $E_{11}$ the above transformations for $g_0\in E_{11}$ can be written as 
$$
g_l\to g_0 g_lg_0^{-1}, \quad g_E\to g_0 g_E\quad {\rm and } \quad g_E\to g_E h
\eqno(1.3)$$
\par
The dynamics of the non-linear realisation is just an action, or set of equations of motion, that are invariant under the transformations of equation (1.2). We  now recall how  to construct the dynamics of the 
the $E_{11}\otimes_s l_1$ non-linear realisation using the  Cartan forms  which are given by 
$$
{\cal V}\equiv g^{-1} d g= {\cal V}_E+{\cal V}_l, 
\eqno(1.4)$$
where 
$$
{\cal V}_E=g_E^{-1}dg_E\equiv dz^\Pi G_{\Pi, \underline \alpha} R^{\underline \alpha}, \ {\rm and }\ 
{\cal V}_l= g_E^{-1}(g_l^{-1}dg_l) g_E= g_E^{-1} dz\cdot l g_E\equiv 
dz^\Pi E_\Pi{}^A l_A  
 \eqno(1.5)$$
Clearly ${\cal V}_E$ belongs to the $E_{11}$ algebra and it is  the Cartan form of $E_{11}$ while ${\cal V}_l$ is in the space of generators of the $l_1$ representation and one can recognise ${ E}_\Pi{}^A = (e^{A_{\underline \alpha}D^{\underline \alpha}})_\Pi{}^A$ as the vielbein on the generalised spacetime. 
\par
Both ${\cal V}_E$ and ${\cal V}_l$ are invariant under rigid transformations,  but  under the local $I_c(E_{11})$ transformations of equation (1.3) they change as 
$$ 
{\cal V}_E\to h^{-1}{\cal V}_E h + h^{-1} d h\quad {\rm and }\quad 
{\cal V}_l\to h^{-1}{\cal V}_l h 
\eqno(1.6)$$


\medskip 
{\bf 2 The eleven dimensional theory}
\medskip
 The theory in eleven dimensions  is found by deleting the  node labelled $11$ of the $E_{11}$ Dynkin diagram and decomposing the $E_{11}\otimes_s l_1$ algebra into representations of the resulting algebra which is GL(11). 
$$
\matrix{
& & & & & & & & & & & & & & \otimes & 11 & & & \cr 
& & & & & & & & & & & & & & | & & & & \cr
\bullet & - & \bullet & - & \bullet & - & \bullet & - & \bullet & - & \bullet & - & \bullet & - & \bullet & - & \bullet & - & \bullet \cr
1 & & 2 & & 3 & & 4 & & 5 & & 6 & & 7 & & 8 & & 9 & & 10 \cr
}
$$
\par
The  $E_{11}$ generators can be classified by a level which  is the number of up minus down indices divided by three. This level is preserved by the $E_{11}$ commutation relations. The decomposition of $E_{11}$ into representations of SL(11) up to level four  can  be found in the book  [6]. 
The positive level generators are [1]
$$
K^a{}_b, \ R^{a_1a_2a_3}, \ R^{a_1a_2\dots a_6} \ {\rm and }\  R^{a_1a_2\ldots a_8,b}, \ldots 
\eqno(2.1)$$
where the generator
$R^{a_1a_2\ldots a_8,b}$ obeys the condition $R^{[a_1a_2\ldots a_8,b
]}=0$ and the indices $a,b,\ldots =1,2\ldots 11$.  
The  negative level generators  are  given by 
$$ 
R_{a_1a_2a_3}, \ R_{a_1a_2\dots a_6} , \ R_{a_1a_2\ldots a_8,b},\ldots
\eqno(2.2)$$
\par
The vector ($l_1$) representation decomposes into representations of $GL(11)$ as [2, 3]
$$
P_a, Z^{ab}, \ Z^{a_1\ldots a_5}, \ Z^{a_1\ldots a_7,b},\  Z^{a_1\ldots a_8},\ 
Z^{b_1 b_2 b_3,a_1 ...a_8}, \ldots  
\eqno(2.3)$$
\par
For the eleven dimensional theory the group element of $E_{11}\otimes_s l_1$ is of the form $g=g_l g_E$ where 
$$
g_E=  \ldots e^{ h_{a_1\ldots a_{8},b}
R^{a_1\ldots a_{8},b}} e^{ A_{a_1\ldots
a_6} R^{a_1\ldots a_6}}e^{ A_{a_1 a_2 a_3} R^{a_1 a_2
a_3}} e^{h_a{}^b K^a{}_b}
\eqno(2.4)$$ 
and 
$$
g_l= e^{x^aP_a} e^{x_{ab}Z^{ab}} e^{x_{a_1\ldots a_5}Z^{a_1\ldots a_5}}\ldots = 
e^{z^A L_A} 
\eqno(2.5)$$
The parameters of the group element $g_E$ will become the fields of the theory while the parameters of the group element $g_l$ will become the  coordinates of the generalised spacetime on which the fields depend. The above parameterisation differs slightly from that used in reference [3] and this will lead to corresponding differences in some of the later  equations in this paper. 
\par
The Cartan forms of $E_{11}$ were defined in equation (1.4) and those of the $E_{11}$ part can be written in the form 
$$
{\cal V} _E= 
G _{a}{}^b K^a{}_b+  G_{c_1\ldots c_3}
R^{c_1\ldots c_3} +G_{c_1\ldots c_6} R^{c_1\ldots c_6}+
G_{c_1\ldots c_8,b} R^{c_1\ldots c_8,b}+\ldots 
\eqno(2.6)$$
\par
We now evaluate these $E_{11}$  Cartan form in terms of the field that parameterise the group element of  equation (2.4),  one finds that  [3]
$$
G _{a}{}^b=(e^{-1}d e)_a{}^b,\ \ G_{a_1\ldots a_3}= e_{a_1}{}^{\mu_1}\ldots e_{a_3}{}^{\mu_3}
dA_{\mu_1\ldots \mu_3}, 
$$
$$ 
G_{a_1\ldots a_6}=  e_{a_1}{}^{\mu_1}\ldots e_{a_6}{}^{\mu_6}(d A_{\mu_1\ldots \mu_6} 
- A_{[ \mu_1\ldots \mu_3}d A_{\mu_4\ldots \mu_6]})
$$
$$
G_{a_1\ldots a_8,b} =     e_{a_1}{}^{\mu_1}\ldots e_{a_8}{}^{\mu_8}e_{b}{}^{\nu}   (d h_{\mu_1\ldots \mu_8,\nu}
-A_{[\mu_1\ldots \mu_3}d A_{\mu_4\mu_5\mu_6} A_{\mu_7\mu_8 ]\nu}
+3 A_{[\mu_1\ldots \mu_6}d A_{\mu_7\mu_8 ]\nu}
$$
$$
+A_{[\mu_1\ldots \mu_3} d A_{\mu_4\mu_5\mu_6} A_{\mu_7\mu_8  \nu]}
-3 A_{[\mu_1\ldots \mu_6}d A_{\mu_7\mu_8 \nu]})
\eqno(2.7)$$
 where $e_\mu{}^a \equiv (e^h)_\mu{}^a$.  
 \par
The generalised vielbein  ${ E}_\Pi{}^A$, can be evaluated from its definition of equation (1.5) to be given as   a matrix  by [3,7]
$$
{ E}= (det e)^{-{1\over 2}}
\left(\matrix {e_\mu{}^a&-3 e_\mu{}^c A_{cb_1b_2}& 3 e_\mu{}^c A_{cb_1\ldots b_5}+{3\over 2} e_\mu{}^c A_{[b_1b_2b_3}A_{|c|b_4b_5]}\cr
0&(e^{-1})_{[b_1}{}^{\mu_1} (e^{-1})_{b_2]}{}^{\mu_2}&- A_{[b_1b_2b_3 }(e^{-1})_{b_4}{}^{\mu_1} (e^{-1})_{b_5 ]}{}^{\mu_2}  \cr
0&0& (e^{-1})_{[b_1}{}^{\mu_1} \ldots (e^{-1})_{b_5]}{}^{\mu_5}\cr}\right)
\eqno(2.8)$$
\par

\medskip
{\bf 3 The transformations of the Cartan forms } 
\medskip

The  Cartan forms are inert under the rigid transformations of equation (1.2)  but under the local Cartan invariant involution transformation  $h \in I_c(E_{11})$   they transform as in equation (1.6). As the Cartan involution invariant subalgebra of SL(11) is SO(11) they transform under SO(11) for the  lowest level transformations. At the next level they transform under a group element  $h$ which  involves the 
 generators at levels $\pm 1$ and it  is of the form 
$$
h=1- \Lambda _{a_1 a_2  a_3 }S^{a_1 a_2  a_3  }, \quad {\rm  where   }\quad
S^{a_1a_2a_3}= R^{a_1a_2a_3}- \eta^{a_1b_1} \eta^{a_2b_2}\ \eta^{a_3b_3} R_{b_1b_2b_3}
\eqno(3.1)$$
Under this transformation the Cartan forms of equation (1.6) change as     
$$
\delta\,{\cal V}_E = \left[S^{a_1a_2a_3}\,\Lambda_{a_1a_2a_3},{\cal V}_E\right] - S^{a_1a_2a_3}\,d\Lambda_{a_1a_2a_3}.
\eqno(3.2)$$ 
These local   variations of the Cartan forms are straightforward to compute, using the $E_{11}$ algebra  and they are given by  [3]
$$
\delta G_{a}{}^{b}=18 \Lambda^{c_1c_2 b }G_{c_1c_2 a}
-2 \delta_a ^{b}  \Lambda^{c_1c_2 c_3}G_{c_1c_2 c_3},\ 
\eqno(3.3)$$
$$
\delta G_{a_1a_2a_3}=-{5!\over 2} G_{b_1b_2b_3 a_1a_2a_3}
\Lambda^{b_1b_2 b_3}-3G^{c}{}_{[a_1 } \Lambda_{|c|a_2a_3]}, -d \Lambda_{a_1a_2a_3}
\eqno(3.4)$$
$$
\delta G_{a_1\ldots a_6}=2 \Lambda_{[ a_1a_2a_3}G_{a_4a_5a_6 ]}
-8.7.2 G_{b_1b_2b_3 [ a_1\ldots a_5,a_6]}\Lambda^{b_1b_2b_3}
+8.7.2 G_{b_1b_2[ a_1\ldots a_5a_6, b_3 ]}\Lambda^{b_1b_2b_3}
\eqno(3.5)$$
$$
\delta G_{a_1\ldots a_8,b}=-3 G_{[ a_1\ldots a_6}\Lambda_{a_7a_8] b}
+3 G_{[ a_1\ldots a_6}\Lambda_{a_7a_8 b]}
\eqno(3.6)$$
\par
When carrying out the local $I_c(E_{11})$ transformations one must take into account the fact that we have also used the local symmetry to fix the group element to have a simpler form,  as we have done in equation (2.4). In most past applications this matter has usually been resolved  by computing 
the compensating local subgroup transformation $h$   required for a given rigid transformation $g_0$ to restore the group element into the chosen form. This involves manipulating group elements and is often very long and cumbersome. In this paper we will use an alternative approach  which was used in  the calculations of reference [5]. The new method  applies to any non-linear realisation for which the local subgroup is the Cartan involution invariant subalgebra, however, to be concrete we will explain it for the case of interest to us here, that is, the $E_{11}\otimes_s l_1$ non-linear realisation. 
\par 
The non-linear realisation $E_{11}\otimes_s l_1$  has a group element $g=g_lg_E$ that is subject to the two types of transformations of equation (1.2) which are required to be symmetries of the dynamics. It is often very useful to use the local transformation to gauge away some of the fields in the group element $g_E$. When the local subgroup is the Cartan involution invariant sub algebra, $I_c(E_{11})$,  we can use the local symmetry to gauge away all the fields associate with negative root generators. In fact it is desirable to keep the level zero symmetries, such as Lorentz symmetry,  manifest and so we only remove all the fields associated with the negative roots except for those at level zero. Put another way we use  the gauge transformations to remove all the negative level fields  from the group element and then the  only remaining local symmetries are those of level zero. We now assume we have made such a choice of group element $g_E$. 
\par
As the group element has only positive level fields and generators,  it follows that the Cartan forms constructed from it will contain only positive level generators.  Although the Cartan forms are  inert under the rigid transformations,  their form is not preserve by the local transformations of equation (1.2), other than by the  transformations of   level zero.  The local transformations which involves level plus and minus one level generators  can be written in the form  $h=1-\Lambda\cdot (R^{(1)}- R^{(-1)})$ and this will not  preserve the form of the Cartan form.  The precise form of this transformation is given in equation (3.1) for the case of eleven dimensions. Such a transformation will result in a change in the Cartan forms that has  a level minus one contribution. To preserve the form of the Cartan forms  we set this contribution to zero and so find the equation
 $$
[\Lambda\cdot R^{(-1)}, {\cal V}^{{(0)} }] - d \Lambda\cdot R^{(-1)}=0
\eqno(3.7)$$
where the superscript denotes the level. 
\par
Equation (3.7) should be thought of as a constraint on the spacetime dependent parameter $\Lambda$  and it can be solved by taking 
$$
\Lambda\cdot R^{(-1)}= (g_E^{(0)})^{-1}\Lambda_c \cdot R^{(-1)}g_E^{(0)}
\eqno(3.8)$$
where $\Lambda_c$ is a constant parameter.  
\par
For the case of eleven dimensions $g_E^{(0)}= e^{h_a{}^b K_a{}^b}$ and we find that equation (3.7) takes the form 
$$
d\Lambda ^{a_1a_2a_3} -3 G_{b}{}^{[a_1 |} \Lambda  ^{ b |a_2a_3]}=0
\eqno(3.9)$$
The solution to this equation is given by 
$$
\Lambda ^{a_1a_2a_3} = \Lambda_c ^{\tau_1\tau_2\tau_3} e_{\tau_1}{}^{a_1} e_{\tau_2}{}^{a_2}
e_{\tau_3}{}^{a_3}
\eqno(3.10)$$
where $\Lambda_c ^{\tau_1\tau_2\tau_3}$ is a constant. The reader may verify that this is the same result as solving equation (3.8). We note that the local transformation is really only a rigid transformation as should indeed be the case as we have fixed the form of the group element using the local transformation. 
\par
We can use equation (3.9) to eliminate $d\Lambda ^{a_1a_2a_3} $ from the transformations of equations (3.3, 3.5, 3.6). In fact this only affects the transformation of the Cartan form for the three form of equation (3.4) which now becomes 
$$
\delta G_{a_1a_2a_3}=-{5!\over 2} G_{b_1b_2b_3 a_1a_2a_3}
\Lambda^{b_1b_2 b_3} -6G_{(c [a_1 |) } \Lambda_{c}{}_{|a_2a_3]}
\eqno(3.11)$$
\par
In carrying out the variation of the equations of motion we will encounter the derivative of the parameter of equation (3.10) and in processing these  terms we must take account of the fact that only $\Lambda_c ^{\tau_1\tau_2\tau_3}$ is independent of the generalised spacetime. This observation plays a crucial role in the calculations in this paper and those of reference [5]. 
\par
The above method is equivalent to carrying out a rigid $E_{11}$ transformation and finding the compensating local transformation that preserves the form of the group element. However, it is very much simpler and easier to implement than the old method. 
\par
The Cartan  forms, discussed above, were written as forms and so  they are  written as $G_{ \underline \alpha}$ where $G_{ \underline \alpha}\equiv dz^\Pi G_{\Pi , \underline \alpha}$ and $G_{\Pi , \underline \alpha}$ are the components. The first index  $\Pi$ is associated with the vector representation  ($l_1$) while the second index is associated with  $E_{11}$. Although the Cartan forms when written in form notation are invariant under the rigid transformations of equation (1.2) once written in terms of components they are not invariant. We can remedy this by taking the first index to be a 
 tangent index, that is, $G_{A , \underline \alpha} = (E^{-1})_A{}^\Pi G_{\Pi , \underline \alpha}$ which is inert under the rigid $E_{11}$ transformations,  but  transforms under the local $I_c(E_{11})$ transformations. This latter   transformation  just being that for the inverse vielbein of equation (1.6). One finds that the Cartan forms, when referred to the tangent space,  transform  on their $l_1$ index as [3]
$$
\delta G_{a, \bullet}= -3G^{b_1b_2}{}_{,\bullet}\ 
\Lambda_{b_1b_2 a},
\quad \delta G^{a_1a_2}{}_{, \bullet}= 6\Lambda^{a_1a_2
b}  G_{b,}{}_{\bullet}
\eqno(3.12)$$
These transformations are to be combined with the local transformations on the second $E_{11}$ index given earlier in this section. 
\medskip 
{\bf 4 Eleven dimensional equations of motion }
\medskip

The non-linear realisation of $E_{11}\otimes_s l_1$ was computed at low  levels  in [3] where one  found  the equation of motion that relates the three form and six form fields. It will be instructive to rederive this equation so as to make clear the origin of terms in the equations of motion that contain derivatives with respect to the higher level coordinates.  On grounds of Lorentz invariance the only equation which is first order in the Cartan forms for the three and six form and has four Lorentz indices must be of the generic form
$$
 { G}_{[a_1,a_2a_3a_4] }-c\, \epsilon _{a_1a_2a_3a_4}{}^{b_1\ldots b_7} G_{[b_1,b_2\ldots b_7] } =0
\eqno(4.1)$$
where $c$ is a constant. We note that the Cartan forms appear with all their indices totally antisymmetrised. 
We now consider the variation of this equation under the transformations of equation (3.1), or equivalently equations (3.3, 3.5, 3.6) and equation  (3.11),   keeping   only the terms that involve the three form and six form.
The variation of ${ G}_{[a_1,a_2a_3a_4] }$ of equation (311) leads to the Cartan form for the six form but it is not totally antisymmetrised in all its indices. Clearly we can only obtain an invariant equation if we have such a total antisymmetry. The way to resolve  this problem is to   consider the object 
$$
{\cal G}_{a_1,a_2a_3a_4 }\equiv  G_{[a_1,a_2a_3a_4] }+{15\over 2}G^{b_1b_2}{}_{, b_1b_2 a_1\ldots a_4}
\eqno(4.2)$$
The variation of this object can be written in the form  
$$
\delta {\cal G}_{a_1,a_2a_3a_4 }= -{1\over 2.4!.4!}\epsilon_{ a_1,a_2a_3a_4 b_1b_2b_3 c_1\ldots c_4} \epsilon ^{  c_1\ldots c_4 e_1\ldots e_7} G_{e_1\ldots e_7} \Lambda ^{b_1b_2b_3}
\eqno(4.3)$$
It is then straightforward to find that, up to the level we are working,  the invariant equation is given by [3]
$$
E_{a_1\ldots a_4}\equiv {\cal G}_{[a_1,a_2a_3a_4] }-{1\over 2.4!}\epsilon _{a_1a_2a_3a_4}{}^{b_1\ldots b_7} G_{b_1,b_2\ldots b_7 }=0
\eqno(4.4)$$
its variation being given by 
 $$
\delta  E_{a_1\ldots a_4}= {1\over 4!} \epsilon _{a_1\ldots a_4 }{}^{b_1\ldots b_7} \Lambda_{b_1 b_2b_3} E_{b_4 \ldots b_7}+\ldots 
\eqno(4.5)$$
where $+\ldots$ denote gravity and higher level contributions. 
\par
Rather than vary the three form equation (4.4) to find the gravity equation we will take the derivative of this equation in such a way as to eliminate the dual six form gauge field and then vary this equation to find the gravity equation. The variation of the first order equation has been given in a previous paper [3],  but its  unfamiliar form and  derivation have  meant that it has not been properly evaluated. 
\par
Taking the derivative of equation (4.4) we find the result 
 $$
\partial_{\nu}( (\det e)^{{1\over 2}} G^{[\nu,\mu_1\mu_2\mu_3]})+
{1\over 2.4!} (\det e)^{{-1}}\epsilon ^{\mu_1\mu_2\mu_3\tau_1\ldots\tau_8} G_{[\tau_1,\tau_2\tau_3\tau_4]} G_{[\tau_5,\tau_6\tau_7\tau_8] }=0 
\eqno(4.6)$$
which is the familiar second order equation of motion for the three form. We have discarded the terms which contain derivatives with respect to the higher level generalised coordinates as we will recover such terms when we vary equation (4.6). When we converted the first ($l_1$) index on the Cartan form to be a tangent index we used the inverse vielbein computed from  the vielbein given in equation (2.8). We notice that the inverse vielbein  contains a factor of $(\det e)^{{1\over 2}}$ compared to what one might normally expect. This explains the unusual factors of $(\det e)^{{1\over 2}}$ that populate the following equations. 
\par
In order to  vary equation (4.6) under the $I_c(E_{11})$ transformation   it is best to rewrite it   in terms of the  Cartan form of $E_{11}$  using the expressions of equation (2.7). We   find that it is equivalent to the equation 
$$
E^{a_1a_2a_3}\equiv E^{a_1a_2a_3 (1)}+E^{a_1a_2a_3(2)}
$$
$$
\equiv  {1\over 2}  G_{b,d}{}^{d} G^{[b, a_1a_2a_3]}- 3G_{b,d}{}^{[a_1|} G^{[b, d |a_2a_3]]}
-G_{c,b}{}^{c} G^{[b, a_1a_2a_3]}+ (\det e)^{{1\over 2}}  e_b{}^\mu\partial_\mu G^{[b, a_1a_2a_3]}
$$
$$
+{1\over 2.4!}\epsilon ^{a_1a_2a_3b_1\ldots b_8} G_{[b_1,b_2 b_3 b_4]} G_{[b_5, b_6 b_7b_8] }=0
\eqno(4.7)$$
The expression $E^{a_1a_2a_3 (1)}$ contains all terms that do not involve the epsilon symbol while  $E^{a_1a_2a_3(2)}$ involves the one term that does. 
\par
We will vary the equations of motion so as to keep in the variations 
all terms that contain ordinary spacetime derivatives. This ensures that we will find all such terms  in the equations of motion. However,  we must also find all terms in the equations of motion we are varying   that contain derivatives with respect to the level one generalised space-time. Indeed,  if we have a term in the variation of  the form 
$$
\Lambda^{\tau \mu_1 \mu_2} G_{\tau, \bullet} f^{\bullet}{}_{\mu_1\mu_2}
\eqno(4.8)$$
then using equation (3.12) we can cancel this  by adding the term 
$$
-{1\over 6}G^{\mu_1\mu_2}{}_{, \bullet} f^{\bullet}{}_{\mu_1\mu_2}
\eqno(4.9)$$
to the equation of motion that is being varied. We will refer to such terms as $l_1$ terms. The other variations of this 
term are of a higher level than we are keeping. Hence when varying a given equation of motion we will find $l_1$ terms in this equation, but not in  the new equation that results from the variation. The latter are then found by varying the new equation. The equations in this paper are given with the understanding that they have been computed up to these levels in the derivatives with respect to the generalised coordinates. 
Of course one can only add  an $l_1$ term to an equation if it has the required index structure and symmetries. 
\par
So that the reader can get a feel for the intricate way in which the calculation works we  now give some indications of how the variations of equation (4.7) under the the local $I_c(E_{11})$ transformation of equation (3.3, 3.5, 3.6) and equation  (3.11) are carried out. 
Varying the  Cartan form $G^{[a_1, a_2 a_3 a_4]}$ contained in $E^{a_1a_2a_3 (1)}$ under the $I_c(E_{11})$ transformation of equation (3.11) and converting the result  back to carry world indices we find the expression 
$$ e_{\mu_1}{}^{[ a_1}e_{\mu_2}{}^{a_2} e_{\mu_3}{}^{a_3]}\{
3\partial_\nu\left( (det\,e )\,\,\omega_{\tau,}{}^{[\nu \mu_1 |}- 
(det\,e )^{{1\over 2}} G_{\tau,}{}^{[\nu \mu_1 |} \, \right)  \Lambda ^{\tau | \mu_2\mu_3]} 
$$
$$
+ {5!\over 2}(\det e)^{{1\over 2}} G^{[\nu, \mu_1\mu_2\mu_3]} {}_{\tau_1\tau_2\tau_3}\Lambda^{\tau_1\tau_2\tau_3}  \} 
\eqno(4.10)$$
When carrying out the variation it is important to recall the discussion of section three and, in particular, the fact that only the parameter $\Lambda ^{\mu_1\mu_2\mu_3}$ is a constant. 
\par
By undoing the  antisymmetrisation  of the four indices we can rewrite the first term as 
$$3 e_{\mu_1}{}^{[ a_1}e_{\mu_2}{}^{a_2} e_{\mu_3}{}^{a_3]}
\partial_\nu\left(det\,e\,\,\omega_{\tau,}{}^{[\nu \mu_1 |}\,\right)  \Lambda ^{\tau | \mu_2\mu_3]}
$$
$$
= {3\over 2} e_{\mu_1}{}^{[ a_1}e_{\mu_2}{}^{a_2} e_{\mu_3}{}^{a_3]}\{ 
\partial_\nu\left(det\,e\,\omega_{\tau,}{}^{\nu \mu_1 }\right) \Lambda ^{\tau \mu_2\mu_3} + \partial_\nu\left(det\,e\,\,\omega_{\tau,}{}^{\mu_1 \mu_2} \right)\Lambda ^{\tau \nu\mu_4} \}
\eqno(4.11)$$
\par
In order to process the first term of equation (4.11) we note that 
$$
e_\mu{}^a\,\partial_\nu\left(det\,e\,\,\omega_{\tau,}{}^{\nu\mu}\,\right) 
$$
$$
= det\,e\,\left(e_b{}^\nu\,\partial_\nu\,\omega_{\tau,}{}^{ba} + \left(e_\mu{}^a\,\partial_\nu\,e_c{}^\mu\right)\,\omega_{\tau,}{}^{\nu c} + \left(e_c{}^\lambda\,\partial_\nu\,e_\lambda{}^c\right)\,\omega_{\tau,}{}^{\nu a} + \partial_\nu\,e_b{}^\nu\,\omega_{\tau,}{}^{ba}\right) ,
\eqno(4.12)$$
the relations 
$$
\left(det\,e\right)\,\omega_{\mu,}{}^{ab}\,\omega_{\lambda,}{}^{c\lambda} = -\omega_{\mu,}{}^{ab}\,\partial_\lambda\left(\left(det\,e\right)\,e_b{}^\lambda\right)
\eqno(4.13)$$
and that 
$$
-\,\omega_{\nu,}{}^a{}_c\,\omega_{\mu,}{}^{c\nu} = \left(G_{a,\,(c\nu)} - G_{c,\,(a\nu)} - G_{\nu,\,[ac]}\right)\,\omega_{\mu,}{}^{c\nu} = -\,G_{c,\,\nu a}\,\omega_{\mu,}{}^{c\nu} = -\,e_b{}^\lambda\,\partial_\nu\,e_\lambda{}^a\,\omega_{\mu,}{}^{\nu b}
\eqno(4.14)$$
\par
The Ricci tensor is given by 
$$
R_\mu{}^a = \partial_\mu\,\omega_{\nu,}{}^{ab}\,e_b{}^\nu - \partial_\nu\,\omega_{\mu,}{}^{ab}\,e_b{}^\nu + \omega_{\mu,}{}^a{}_c\,\omega_{\nu,}{}^{cb}\,e_b{}^\nu - \omega_{\nu,}{}^a{}_c\,\omega_{\mu,}{}^{cb}e_b{}^\nu.
\eqno(4.15)$$
whereupon we  recognise that the frist term in equation (4.11) is just the first, third and fourth terms of the Ricci tensor and as a result  we can write this term  as 
$$
{3\over 2} \det e\{ R_\tau {}^{[ a_1 |}- 
\partial_\tau (\omega_\nu, {}^{[ a_1 | b}) e_b{}^\nu  \} \Lambda ^{\tau |a_1a_2]} 
\eqno(4.16)$$
However, the second term in equation (4.11) is of the form of equation (4.9) and so it can be introduced by adding an $l_1$ term to the three form equation of motion.  Of course we can not from these considerations determine the coefficient of this term to be exactly as required to find the full Ricci tensor. However, this coefficient is fixed to the desired result once we vary the resulting gravity equation as was shown at the linearised level in [5]. For simplicity of presentation we will take the coefficient to be as required. The reader who wishes to insert an arbitrary coefficient and follow it through the remaining calculations, including the non-linear variation of the gravity equation,  is encouraged to do so.  
\par
The second terms in both equations (4.10) and (4.11)  are of the form $G_{\tau ,\bullet }\Lambda^ {\tau \mu\nu}$ and so they    can  be cancelled by adding  $l_1$ terms  to the three form  equation of motion. 
\par
The variation of the second term in equation (4.7), that is,  the terms in $E^{a_1a_2a_3 (2)}$ can be processed by using equation (4.4) 
to swop the seven form field strength for the four form  field strength. One finds the terms associated with the energy momentum tensor, further $l_1$ terms and terms which are cancelled by the variation of $E^{a_1a_2a_3 (1)}$. 
\par
 As explained,   above when carrying out the variation of the three form equation we find the $l_1$ terms that we must add to this equation. The result of  all these calculations is that the three form equation of motion, up to the level  we are calculating,  now takes the form 
$$
{\cal E} ^ {a_1a_2a_3}\equiv {1\over 2}  G_{b,d}{}^{d} G^{[b, a_1a_2a_3]}- 3G_{b,d}{}^{[a_1|} G^{[b, d |a_2a_3]]}
-G_{c,b}{}^{c} G^{[b, a_1a_2a_3]}+ (\det e)^{{1\over 2}}e_b{}^\mu\partial_\mu G^{[b, a_1a_2a_3]}
$$
$$
+{1\over 2.4!}\epsilon ^{a_1a_2a_3b_1\ldots b_8} G_{[b_1,b_2 b_3 b_4]} G_{[b_5, b_6 b_7b_8] }
$$
$$  -\,9\,G^{ca_1}{}_{,cd_1d_2}\,G^{[d_1,\,d_2a_2a_3]} + {5\over 16}\,\varepsilon^{a_1a_2a_3b_1...b_8}\,G_{b_1,\,b_2b_3b_4}\,G^{c_1c_2}{}_{,c_1c_2b_5...b_8}
$$
$$
+ {1\over 4}\,e_{\mu_1}{}^{[a_1}e_{\mu_2}{}^{a_2}e_{\mu_3}{}^{a_3 ]}
\partial_\nu \left( (\det e)^{{1\over 2}} G^{\mu_1\mu_2}{}_{,}{}^{\nu\mu_3 }\right)
+{1\over 4} (\det e)^{{1\over 2}} \omega_{\nu,} {}^{[a_1 |b} G^{a_2a_3 ]}{}_{, b}{}^{\nu}
$$
$$
+{1\over 4} G^{[a_1a_2 |}{}_{, d} {}^{d} (G^{|a_3 ] }{}_{, c}{}^{c}-G_{c,}{}^{ |a_3 ]c})
+{1\over 4} \partial_\nu \left( (\det e)^{{1\over 2}}  (G^{[ a_1a_2 |}{}_{,d}{}^{d}e^{\nu |a_3]}-
G^{[ a_1a_2 |}{}_{,}{}^{ |a_3 ]\nu} )\right)
$$
$$
+{1\over 2} ( G^{[a_1}{}_{,c}{}^{a_2} G^{|c | a_3]}{}_{,d}{}^{d} -G^{c [a_1}{}_{,} {}^{a_2 |e|} G_{e,c}{}^{a_3]})
$$
$$
+{15\over 2} e_{\mu_1}{}^{[a_1}e_{\mu_2}{}^{a_2}e_{\mu_3}{}^{a_3 ]}
\partial_\nu\left((\det e)^{{1\over 2}} G^{d_1d_2}{}_{, d_1d_2} {}^ {\nu \mu_1\mu_2\mu_3}\right)
$$
$$
+e_{\mu_1}{}^{[a_1}\,e_{\mu_2}{}^{a_2}\,e_{\mu_3}{}^{a_3]}\,\Big({1\over 2}\,\left(det\,e\right)^{1\over 2}\,g_{\tau\sigma}\,g^{\mu_1\lambda}\,\partial_{\lambda}\,G^{\tau\mu_2,\,\left(\sigma\mu_3\right)} - {1\over 2}\,G^{\tau\mu_1}{}_{,\,d}{}^d\,G^{\mu_2,\,(\mu_3}{}_{\tau)} 
$$
$$
- {1\over 4}\,G^{\tau\mu_1,\,(\mu_2}{}_{\tau)}\,G^{\mu_3}{}_{,\,d}{}^d-\,G^{\tau\mu_1}{}_{,\,(\tau\sigma)}\,G^{\mu_2,\,(\mu_3\sigma)} + G^{\tau\mu_1,\,(\mu_2\sigma)}\,G_{\sigma,\,(\tau}{}^{\mu_3)}\Big)=0
\eqno(4.17)$$
\par
Under the variation of the local transformations of equation (3.1) this equation of motion  transforms as  
$$
\delta {\cal E}^{a_1a_2a_3 }= {3\over 2} E_b{}^{[a_1|}\Lambda ^{b | a_2a_3]} 
$$
$$
+{1\over 24} e_{\mu_1}^{a_1}e_{\mu_2}^{a_2} e_{\mu_3}^{a_3} \epsilon ^{\mu_1\mu_2\mu_3 \nu\lambda_1\ldots \lambda_4 \tau_1\tau_2\tau_3}\partial_\nu \left( (\det e)^{-{1\over 2}} E_{\lambda_1\ldots \lambda_4}g_{\tau_1 \kappa_1} g_{\tau_2 \kappa_2}
g_{\tau_3 \kappa_3}\right) \Lambda^{ \kappa_1\kappa_2\kappa_3}
$$
$$
+{1\over 24.4!} \epsilon^{a_1a_2a_3b_1\ldots b_8} \epsilon _{b_1\ldots b_4 c_1c_2c_3 e_1\ldots e_4} E_{b_5\ldots b_8} G^{[e_1,e_2\ldots e_4 ]} \Lambda ^{c_1c_2c_3}
\eqno(4.18)$$
where 
$$
E_a{}^b\equiv (\det e)  R_a{}^b- 12.4 G_{[a, c_1c_2c_3]}G^{[b, c_1c_2c_3]}+4\delta _a^b G_{[c_1, c_2c_3c_4]}G^{[c_1, c_2c_3c_4]}=0
\eqno(4.19)$$
and $E_{\lambda_1\ldots \lambda_4}$ is the the first order in derivatives duality relation of equation (4.4). 
Clearly, the graviton equation of motion is equation (4.19). 
\par
We will now carry out the variation of the gravity equation under the $I_c(E_{11})$ transformation of equation (3.3, 3.5, 3.6) and equation  (3.11).  This calculation requires  the variation of the spin connection which we have defined  to be given by 
$$
(\det e)^{{1\over 2}} \omega _{c, ab}= - G_{a, (bc)}+ G_{b, (ac)}+G_{c, [ab]}
\eqno(4.20)$$
Since the $G_{A, \underline \alpha}$ contain a factor  of $(\det e)^{{1\over 2}}$ this is the standard expression for the spin connection. 
 The variation  will  result in only  the four form field strength $G_{[c_1,c_2c_3c_4 ]}$ if we add to the spin connection certain $l_1$ terms. Indeed if we define   
$$
(\det e)^{{1\over 2}}\Omega _{c, ab}= (\det e)^{{1\over 2}}\omega_{c, ab}
-3 G^{dc}{}_{, dab} -3 G^{d}{}_{b}{}_{, dac} +3 G^{d}{}_{a}{}_{, dbc} 
-\eta _{bc} G^{d_1d_2}{}_{, d_1d_2 a}+\eta _{ac} G^{d_1d_2}{}_{, d_1d_2 b}
\eqno(4.21)$$
one then finds that 
$$
\delta ((\det e)^{{1\over 2}}\Omega _{c, ab})=-18.2 \Lambda ^{ d_1d_2}{}_{ c} G_{[a,bd_1d_2]} -18.2 \Lambda ^{d_1d_2}{}_{ b} G_{[a,c d_1d_2 ]}-18.2 \Lambda ^{d_1d_2}{}_{ a} G_{[c,b d_1d_2]}
$$
$$
+8 \eta _{bc}  \Lambda ^{d_1d_2 d_3} G_{[a, d_1d_2d_3 ]}-8 \eta _{ac}  \Lambda ^{d_1d_2 d_3} G_{[b, d_1d_2d_3 ]}
\eqno(4.22)$$
\par
Substituting the spin connection $\Omega _{c, ab}$ for the standard spin connection $\omega _{c, ab}$ in the Riemann tensor we define 
$$
{\cal R}_a{}^b = e_a{}^\mu \partial_\mu\,\Omega_{\nu,}{}^{bd}\,e_d{}^\nu - e_a{}^\mu \partial_\nu\,\Omega_{\mu,}{}^{bd}\,e_d{}^\nu + \Omega_{a,}{}^b{}_c\,\Omega_{d,}{}^{cd} - \Omega_{d,}{}^b{}_c\,\Omega_{a,}{}^{cd}. 
\eqno(4.23)$$
In fact ${\cal R}_a{}^b$ is no longer symmetric in $a$ and $b$ interchange 
when we consider the terms that have level one derivatives in the generalised coordinates. We replace the Ricci tensor by the object of equation (4.23) in the  equation of motion of equation (4.19). We will then require its variation which is given by  
$$
\delta \{ (\det e) {\cal R}_{ab}\}= \{ 36\Lambda ^{d_1d_2 c} (\omega_{c, a}{}^{e}+\omega^{e}{}_{, a c})G_{[b, ed_1d_2]}
-36 \Lambda^{d_1d_2}{}_{ b}\  \omega^{c,}{}_{ a}{}^{e} G_{[c, ed_1d_2]}
$$
$$
+ 18 G^{e}{}_{,c}{}^{c} \Lambda ^{d_1d_2}{}_{ b} G_{[a, ed_1d_2]} 
-36 G_{c,}{}^{e c} \Lambda ^{d_1d_2}{}_{ b} G_{[a, ed_1d_2]} 
+(a\leftrightarrow b) \}
$$
$$
+\eta_{ab} (-8G_{e,}{}^{ce}+4 G^{c}{}_{, e}{}^{e })\Lambda ^{d_1d_2 d_3}G_{[c,  d_1d_2d_3] }
-18 G_{e, c}{}^{c} \Lambda ^{d_1d_2e} G_{[a, b d_1d_2]} 
$$
$$
+36 G_{c,e}{}^{c} \Lambda ^{d_1d_2 e} G_{[a, b d_1d_2]} 
+ 18\omega _{a, b} {}^{e}  \Lambda ^{d_1d_2 c}G_{c, e d_1d_2} 
+6 \omega _{e, b } {}^{e}\Lambda ^{d_1d_2 d_3}G_{ d_1 , d_2d_3 a}
\eqno(4.24)$$

$$
+\,8\,\left(det\,e\right)^{1\over 2}\,e_a{}^\mu\,\partial_\mu\,\left(\,G_{[b,\,\tau_1\tau_2\tau_3]}\,\Lambda^{\tau_1\tau_2\tau_3}\right)
$$
$$
+\,\left(det\,e\right)^{1\over 2}\,e^c{}^\nu\,\partial_\nu\,\Big(36\,\Lambda^{d_1d_2}{}_{a}\,G_{[b,\,cd_1d_2]} + 36\Lambda^{d_1d_2}{}_{c}\,G_{[b,\,ad_1d_2]} + 36\Lambda^{d_1d_2}{}_{b}\,G_{[a,\,cd_1d_2]}
$$
$$
-\,8\,\eta_{ac}\,G_{[b,\,d_1d_2d_3]}\,\Lambda^{d_1d_2d_3} + 8\,\eta_{ab}\,G_{[c,\,d_1d_2d_3]}\,\Lambda^{d_1d_2d_3}\Big)
$$
\par
Calculating the other variation of the other terms in the gravity equation of motion (4.19) we find that its variation is given by 
$$
\delta {\cal E}_{ab}= -36 \Lambda ^{d_1d_2 }{}_{a} { E}_{bd_1d_2} -36 \Lambda ^{d_1d_2 }{}_{b} { E}_{ad_1d_2} 
+8\eta_{ab}\Lambda ^{d_1d_2d_3 } E_{ d_1d_2d_3} 
$$
$$
-2 \epsilon_{a}{}^{c_1c_2c_3 e_1\ldots e_4 f_1f_2f_3} \Lambda_{f_1f_2f_3} 
E_{e_1\ldots e_4}G_{[b, c_1c_2c_3]}
-2 \epsilon_{b}{}^{c_1c_2c_3 e_1\ldots e_4 f_1f_2f_3} \Lambda_{f_1f_2f_3} 
E_{e_1\ldots e_4}G_{[a, c_1c_2c_3]}
$$
$$ +{1\over 3} \eta_{ab} \epsilon^{c_1\dots c_4 e_1\ldots e_4 f_1f_2f_3}
E_{e_1\ldots e_4}G_{[c_1, c_2c_3c_4]}\Lambda_{f_1f_2f_3} 
\eqno(4.25)$$
where 
$$
{\cal E}_{a b}\equiv (\det e)  {\cal R}_{ab}- 12.4 G_{[a, c_1c_2c_3]}G^{[e, c_1c_2c_3]}\eta_{eb}+4\eta_{ ab} G_{[c_1, c_2c_3c_4]}G^{[c_1, c_2c_3c_4]}
$$
$$
-\,3.5!\,G^{d_1d_2}{}_{,\,d_1d_2a}{}^{c_1c_2c_3}\,G_{[b,\,c_1c_2c_3]} - 3.5!\,G^{d_1d_2}{}_{,\,d_1d_2b}{}^{c_1c_2c_3}\,G_{[a,\,c_1c_2c_3]} 
$$
$$+ {5!\over 2}\,\eta_{ab}\,G^{d_1d_2}{}_{,\,d_1d_2c_1...c_4}\,G^{[c_1,\,c_2c_3c_4]}
 - 12\,G^{c_1c_2}{}_{,\,a}{}^{c_3}\,G_{[b,\,c_1c_2c_3]} 
+3 G^{c_1c_2}{}_{, e}{}^{e} G_{[a,bc_1c_2]}
$$
$$
-6\left(\det\,e\right)\,e_{b}{}^\mu\,e_a{}^\lambda\,\partial_{[\mu |}\left[\left(det\,e\right)^{-\,{1\over 2}}G^{\tau_1\tau_2}{}_{,\,| \lambda\tau_1\tau_2]}\right]
$$
$$
-\,\left(det\,e\right)^{{1\over 2}}\omega_c{}_{,}{}_{b}{}^{c}\,G^{d_1d_2}{}_{,\,d_1d_2a} - 3\,\left(det\,e\right)^{{1\over 2}}\omega_a{}_{,}{}_{b}{}^{c}\,G^{d_1d_2}{}_{,\,d_1d_2c}
=0
\eqno(4.26)$$
In carrying out the variation we find the $l_1$ terms we must add to the gravity equation which are now included above.  We note that some of the terms in equation (4.26)  are not symmetric under the interchange of  $a$ and $b$  so compensating the same lack of symmetry in ${\cal R}_{ab}$. 
\par
Thus we have found that, up to the level at which we are working, the second order in derivatives three form and gravity equations (4.17) and (4.26) respectively rotate into each other as well as the first order duality equation (4.4). However, once we vary the latter equation we will find equations of motion for the higher level fields in the $E_{11}\otimes _s l_1,$ non-linear realisation and hence the higher order fields can only be eliminated from the complete system by truncating in a way that destroys the $E_{11}$ symmetry. 
\par
We recognise equations (4.17) and equation (4.26) as the equations of motion of the bosonic sector of eleven dimensional supergravity once we throw away the terms that have derivatives with respect to the level one generalised coordinates. 


\medskip
{\bf {5 Conclusion}}
\medskip
In this paper we have constructed the dynamics that follow from the non-linear realisation of $E_{11}\otimes_s l_1$  in eleven   dimensions for the low level fields and generalised coordinates. The result is unique and when we truncate it to contain only the usual fields of supergravity, that is, the graviton and the three form, and also take only the usual coordinates of spacetime we find the equations of motion of eleven dimensional supergravity. Thus we have a very direct path from the Dynkin diagram of $E_{11}$ to the eleven dimensional supergravity theory. It is inevitable that one will find the analogous results in other dimensions. Indeed the five dimensional theory was found in reference [5] except for some   coefficients which were undetermined, however,  these can be fixed to the required values from the eleven dimensional theory using dimensional reduction. 
\par
The $E_{11}\otimes_s l_1$ realisation is a unified theory in that it contains all the maximal supergravities in one theory. The theory in $D$ dimensions appears by deleting node $D$ in the $E_{11}$ Dynkin digram and decomposing $E_{11}\otimes_s l_1$ with respect to the resulting $GL(D)\times E_{11-D}$ algebra [8,9,10]. The  $E_{11}\otimes_s l_1$  also includes the gauged supergravities [9,10,11]. Furthermore,  it  includes effects that are beyond the usual  supergravity description and are know to be present in the theory of strings and branes.  Since the supergravity theories themselves contain many of the low energy properties of strings and branes it would seem inevitable  that one should   replace the many  different  supergravity theories by the $E_{11}\otimes_s l_1$ realisation as the low energy effective theory of strings and branes. 
\par
 The $E_{11}$ theory is very predictive in that one can, at least  as a matter of principle find how the higher level fields and coordinates enter into the equations of motion. It would be very interesting to find what are the physical meaning of the higher level fields and coordinates.  Reference [5] mentions  a number of  avenues  that one can explore in future work.

\medskip
{\bf {Acknowledgements}}
\medskip
We wish to thank Nikolay Gromov for help with the derivation of the equations of motion from the non-linear realisation. We also wish to thank the SFTC for support from Consolidated grant number ST/J002798/1 and Alexander Tumanov wishes to thanks King's College  for the support provided by his  Graduate School International Research Studentship. 
\medskip
{\bf {References}}
\medskip
\item{[1]} P. West, {\it $E_{11}$ and M Theory}, Class. Quant.  
Grav.  {\bf 18}, (2001) 4443, hep-th/ 0104081. 
\item{[2]} P. West, {\it $E_{11}$, SL(32) and Central Charges},
Phys. Lett. {\bf B 575} (2003) 333-342,  hep-th/0307098. 
\item {[3]} P. West, {\it Generalised Geometry, eleven dimensions
and $E_{11}$}, JHEP 1202 (2012) 018, arXiv:1111.1642.  
\item{[4]} P. West, {\it  E11, Generalised space-time and equations of motion in four dimensions}, JHEP 1212 (2012) 068, arXiv:1206.7045. 
\item{[5]} A. Tumanov and P. West, {\it E11 must be a symmetry of strings and branes },  arXiv:1512.01644. 
\item{[6]} P. West, {\it Introduction to Strings and Branes}, Cambridge University Press, 2012. 
\item{[7]} A. Tumanov and P. West, {\it Generalised vielbeins and non-linear realisations },  arXiv:1405.7894. 
\item{[8]} P. West, {\it The IIA, IIB and eleven dimensional theories 
and their common
$E_{11}$ origin}, Nucl. Phys. B693 (2004) 76-102, hep-th/0402140. 
\item{[9]}  F. ÊRiccioni and P. West, {\it
The $E_{11}$ origin of all maximal supergravities}, ÊJHEP {\bf 0707}
(2007) 063; ÊarXiv:0705.0752.
\item{[10]} ÊF. Riccioni and P. West, {\it E(11)-extended spacetime
and gauged supergravities},
JHEP {\bf 0802} (2008) 039, ÊarXiv:0712.1795.
\item{[11]} E. Bergshoeff, I. De Baetselier and  T. Nutma, {\it 
E(11) and the Embedding Tensor},  JHEP 0709 (2007) 047, arXiv:0705.1304. 

\end

The l1 term:
$$
e_{\mu_1}{}^{[a_1}\,e_{\mu_2}{}^{a_2}\,e_{\mu_3}{}^{a_3]}\,\Big({1\over 4}\,\left(det\,e\right)^{1\over 2}\,g^{\mu_1\lambda}\,\left(\partial_\lambda\left(G^{\mu_3}{}_{\tau,}{}^{b\tau}\right)\,e_b{}^{\mu_2} + \partial_\lambda\left(G^{\mu_3}{}_{\tau,}{}^{b\mu_2}\right)\,e_b{}^{\tau}\right)
$$
$$
+\,{1\over 2}\,G^{\mu_1}{}_{\tau,\,d}{}^d\,G^{\mu_2,\,(\mu_3\tau)} + {1\over 4}\,G^{\mu_1}{}_{\tau,}{}^{(\mu_2\tau)}\,G^{\mu_3}{}_{,\,d}{}^d
$$
$$
-\,G^{\mu_1}{}_{\tau,}{}^{(\sigma\mu_2)}\,G_{\sigma,}{}^{(\mu_3\tau)} + G^{\mu_1\sigma}{}_{,\,(\tau\sigma)}\,G^{\mu_2,\,(\mu_2\tau)} + G^{\sigma\mu_1,\,(\mu_2\tau)}\,G^{\mu_3}{}_{,\,(\tau\sigma)} + {1\over 2}\,G^{\mu_1}{}_{\tau,}{}^{b(\mu_2|}\,G^{\mu_3}{}_{,\,b}{}^{|\tau)}\Big).
$$
